\documentstyle[prl,aps,twocolumn]{revtex}
\title{Indistinguishability and nonlocality in Einstein-Podolsky-Rosen experiment}
\author{Adonai S. Sant'Anna}
\address{Dep. Matem\'atica, UFPR, C.P. 019081, Curitiba, PR, 81531-990, Brazil\\e-mail: adonai@scientist.com, Phone: +55-41-345-9413, FAX: +55-41-267-4236}
\begin{document}
\newtheorem{obs}{\bf Remark:}
\newtheorem{definicao}{\bf Definition}
\newtheorem{teorema}{\bf Theorem}
\newtheorem{lema}{\bf Lemma}
\setlength{\unitlength}{1mm}
\renewcommand{\thefootnote}{\arabic{footnote}}
\setcounter{footnote}{0}
\newcounter{cms}
\setlength{\unitlength}{1mm}

\maketitle

\begin{abstract}
Quasi-set theory allows us a non trivial relation between indistinguishability and nonlocality into the context of Einstein-Podolsky-Rosen experiment. Quasi-set theory is a set theory which provides a manner for dealing with collections of indistinguishable but not identical elementary particles.
\end{abstract}
\vskip2mm
\draft{PACS numbers: 02.10.Cz, 02.40.Pc, 03.65.Bz}

\draft{Keywords: indistinguishability, nonlocality, EPR, quasi-sets}

\section{Introduction}

	The recent experiments on teleportation \cite{Bouwmeester-97} demonstrate that instantaneous transportation of information is an experimental fact. The quantum state of a given system may be transported from one location to another without moving through the intervening space. Partial implementations of quantum teleportation over macroscopic distances have been achieved by using optical systems. That suggests, in our opinion, that a revision on the very structure of space-time is demanded. Actually, teleportation is essentially based on Einstein-Podolsky-Rosen (EPR) experiment \cite{Sakurai-94}. We claim that we should revise to notion of space-time at microscopic levels, by taking into account the non-individuality of elementary particles. If two particles are indistinguishable, then we should not ascribe usual space-time coordinates to them, since these coordinates act like labels.

	In this paper we consider that `indistinguishable' objects are objects that share their properties, while `identical objects' means `the very same object', and not two objects at all. We need to settle this vocabulary, since in physics textbooks it is usual to consider the words `indistinguishable' and `identical' as synonymous.

	Relativity does not allow signals with velocities greater than the speed of light in the vacuum. EPR suggests that there is an instantaneous communication between two elementary particles.

	It is considered that the interference produced by two light beams, in a two-slit experiment, is determined by both their mutual coherence and the indistinguishability of the quantum particle paths. For instance, Mandel \cite{Mandel-91} has proposed a quantitative link between the wave and the particle descriptions by using an adequate decomposition of the density operator. In this paper we propose the use of a set-theoretical framework without identity for the EPR {\em Gedanken\/} experiment, which allows us to show that nonlocal phenomena between entangled particles is a logical consequence from their non-individuality.

	The set-theory with no identity that we use is quasi-set theory \cite{Krause-96}. Quasi-set theory is based on Zermelo-Fraenkel axioms for sets and permits to cope with collections of indistinguishable objects by allowing the presence of two sorts of atoms ({\it Urelemente\/}), termed $m$-atoms and $M$-atoms. A binary relation of indistinguishability between $m$-atoms (denoted by the symbol $\equiv$), is used instead of identity, and it is postulated that $\equiv$ has the properties of an equivalence relation. The predicate of equality cannot be applied to the $m$-atoms, since no expression of the form $x = y$ is a well-formed formula if $x$ or $y$ denote $m$-atoms. Hence, there is a precise sense in saying that $m$-atoms can be indistinguishable without being identical.

	We have recently proposed (with D. Krause and A. G. Volkov) that quasi-set theory provides a tool for dealing with collections of indistinguishable elementary particles \cite{Krause-99}. In \cite{Sant'Anna-00} we proved (with A. M. S. Santos) that a Maxwell-Boltzmann distribution is possible even in an ensemble of indistinguishable particles. In this paper we show that indistinguishability implies nonlocality as it follows in the next paragraphs. 

\section{EPRB}

	The original Einstein-Podolsky-Rosen {\em Gedanken\/} experiment dealt with measurements of position and momentum in a two-particle system. Here we appeal to the use of a composite spin $\frac{1}{2}$ system started with D. Bohm. This kind of experimental setup we refer to as Einstein-Podolsky-Rosen-Bohm (EPRB) experiment. Our discussion on this topic is essentially based on the classic textbook by J. J. Sakurai \cite{Sakurai-94}.

	It is well known that the state ket of a two-electron system in a spin-singlet state can be described by:

\begin{equation}
\left(\frac{1}{\sqrt{2}}\right)(|{\bf z}+;{\bf z}-\rangle - |{\bf z}-;{\bf z}+\rangle ),
\end{equation}

\noindent
where ${\bf z}$ is an arbitrary quantization direction, and $|{\bf z}+;{\bf z}-\rangle$ means that electron $1$ is in the spin-up state while electron $2$ is in the spin-down state. Something similar may be said about $|{\bf z}-;{\bf z}+\rangle$. 

	If the spin component of particle $1$ is shown to be, e.g., in the spin-up state, the other particle component is necessarily in the spin-down state. This remarkable correlation has been experimentaly confirmed \cite{Aspect-99}. Some authors have discussed the possibility of a comunication between entangled particles by means of `wormholes' \cite{Holland-95}. In this paper we suggest another topological explanation of this weird phenomenum.

\section{Quasi-Metric Spaces}

	Our mathematical framework for describing EPRB is quasi-sets. Quasi-set theory ${\cal Q}$ allows the presence of two sorts of atoms ({\it Urelemente\/}), termed $m$-atoms and $M$-atoms, identified by two unary predicates $m(x)$ and $M(x)$, respectively. It is important to observe that the term `atom' is in the mathematical sense. Concerning the $m$-atoms, a weaker `relation of indistinguishability' (denoted by the symbol $\equiv$), is used instead of identity, and it is postulated that $\equiv$ has the properties of an equivalence relation.

	The universe of ${\cal Q}$ is composed by $m$-atoms, $M$-atoms and {\it quasi-sets\/}. The sentence `$x$ is a quasi-set' is denoted by $Q(x)$, where $Q$ is a unary predicate. The axiomatics is adapted from that of ZFU (Zermelo-Fraenkel with {\it Urelemente\/}), and when we restrict the theory in not considering $m$-atoms, quasi-set theory is essentially equivalent to ZFU, and the corresponding quasi-sets will be termed `ZFU-sets' (similarly, if also the $m$-atoms are ruled out, the theory collapses into ZFC). The $M$-atoms play the role of the {\it Urelemente\/} in the sense of ZFU.

	In order to preserve the concept of identity for the `well-behaved' objects, an {\it extensional equality\/} is introduced for those entities which are not $m$-atoms on the following grounds: for all $x$ and $y$, if they are not $m$-atoms, then $x =_{E} y$ iff $(Q(x)\wedge Q(y)\wedge(\forall z ( z \in x \Leftrightarrow z \in y ))) \vee (M(x) \wedge M(y) \wedge x \equiv y)$. We are using standard logical notation: $\Rightarrow$ is the conditional of propositional calculus, $\neg$ is negation, $\wedge$ is conjunction, $\vee$ is disjunction, $\Leftrightarrow$ is the biconditional, and $\forall$ and $\exists$ are, respectively, the universal and the existential quantifiers of predicate calculus.

	It is possible to prove that $=_{E}$ has all the properties of classical identity and so these properties hold regarding $M$-atoms and `ZFU-sets'. It is straight to see that we can easily define the binary relations ``$\neq_E$'', ``$<_E$'', ``$>_E$'', ``$\leq_E$'', and ``$\geq_E$'', as natural extensions, respectively, of ``$\neq$'', ``$<$'', ``$>$'', ``$\leq$'', and ``$\geq$'' in the ZFU-set of, e.g., real numbers. 

	According to the weak-pair axiom in quasi-set theory, for all $x$ and $y$, there exists a quasi-set whose elements are the indistinguishable objects from either $x$ or $y$. In symbols: $\forall x \forall y \exists z (Q(z) \wedge \forall t (t \in z \Leftrightarrow t \equiv x \vee t \equiv y))$. Such a quasi-set is denoted by $[x, y]$ and, when $x \equiv y$, we have $[x]$ by definition. We remark that this quasi-set {\em cannot\/} be regarded as the `singleton' of $x$, since its elements are {\em all\/} the objects indistinguishable from $x$, so its `cardinality' (or quasi-cardinality, if we use the original terminology in \cite{Krause-96}) may be greater than $1$. A concept of {\it strong singleton\/}, which plays an important role in the applications of quasi-set theory, may be defined. We call $[x]$ a {\em weak singleton\/}.

	It is rather intuitive that the concept of {\em function\/} cannot be defined in the standard way, so it is introduced a weaker concept of {\em quasi-function\/}, which maps collections of indistinguishable objects into collections of indistinguishable objects. When there are no $m$-atoms involved, the concept is reduced to that of function as usually understood. 

	The concept of {\em relation\/} is like the standard one: a quasi-set $w$ is a relation between two quasi-sets $x$ and $y$ if $w$ satisfies the following predicate $R$: $R(w)$ iff $Q(w) \wedge \forall z (z \in w \Rightarrow \exists u \exists v (u \in x \wedge v \in y \wedge z =_{E} \langle u, v \rangle))$. The notion of ordered pair $\langle u, v \rangle$ is analogous to the usual definition.

	As usual, if $x =_{E} y$, we say that $R$ is a relation {\em on\/} $x$. We denote by $Dom(R)$ (the {\em domain} of $R$) the quasi-set $[u \in x : \langle u, v \rangle \in R ]$ and by $Rang(R)$ (the {\em range} of $R$) the quasi-set $[v \in y : \langle u, v \rangle \in R]$.  

	Let $x$ and $y$ be quasi-sets. We say that $f$ is a {\em quasi-function\/} from $x$ to $y$ if $f$ is such that ($R$ is the predicate for `relation' defined above): $R(f) \wedge \forall u (u \in x \Rightarrow \exists v (v \in y \wedge \langle u, v \rangle \in f)) \wedge \forall u \forall u' \forall v \forall v' (\langle u, v \rangle \in f \wedge \langle u', v'
\rangle \in f \wedge u \equiv u' \Rightarrow v \equiv v')$.

	As a final remark on this review on quasi-sets we say that there is a unary functional letter in the language of quasi-set theory named as $qc$. If $x$ is a variable, then $qc(x)$ corresponds to the {\em quasi-cardinality\/} of $x$, which is a cardinal. Roughly speaking, $qc(x)$ corresponds to the number of elements of $x$ and it is a natural extension of the usual notion of cardinality of a set.

	Next we define metric spaces into the context of quasi-set theory. Our main goal is to define a very simple quasi-set-theoretical structure for a space-time of a system of two EPRB-correlated particles.

\begin{definicao}\label{qmetric}
A quasi-metric space is an ordered pair $\langle X,d\rangle$, such that:

\begin{enumerate}

\item $X$ is a non-empty quasi-set.

\item $d:X\times X\to\Re$ is a quasi-function which associates each ordered pair $\langle x,y\rangle$ to a real number $d(x,y)$, which we refer to as the distance between $x$ and $y$.

\item $d(x,y) =_E 0$ if and only if $x\equiv y$.

\item $d(x,y)>_E 0$ if and only if $\neg(x\equiv y)$, that is, $x$ and $y$ are {\em distinguishable\/} ($\neg$ is the standard negation connective from mathematical logic).

\item $d(x,y) =_E d(y,x)$.

\item $d(x,z) \leq_E d(x,y) + d(y,z)$

\end{enumerate}

\end{definicao}

	The above definition is a very natural generalization of the usual concept of metric space, which certainly deserves an adequate mathematical study that we do not accomplish here, since that task is out of the scope of this paper.

	Now we present an example of a non-trivial quasi-metric space with its correspondent physical interpretation in terms of the EPRB {\em Gedanken\/} experiment.

\begin{definicao}\label{xxxx}
An EPRB-Space is a quasi-metric space ${\cal S} =_E \langle M,d_q\rangle$, where:

\begin{description}

\item[A1] $M =_E V \cup [x]_2$, where $V\subset\Re^n$ is an open set of the usual set of ordered $n$-tuples of real numbers, endowed with the euclidian metric;\footnote{It could be a flat metric of Lorentz signature.} $[x]_2$ is a weak singleton such that $m(x)$; and $qc([x]_2) =_E 2$.

\item[A2] If $a$ and $b$ are elements of $V$ then $d_E(a,b) \leq_E 2c$, where $d_E$ is the euclidian distance, and $c$ is a real constant such that $c >_E 0$.

\item[A3] $d_q(x,a) =_E d_q(a,x) =_E c$ for all $x\in [x]_2$, and for all $a\in V$.

\item[A4] $d_q(a,b) =_E d_E(a,b)$ if $a$ and $b$ are both elements of $V$.

\end{description}
\end{definicao}

	We know that the entanglement state of the two correlated particles in EPRB entails a relation of indistinguishability between them. So, such particles cannot be labeled by anything. {\em The EPRB-correlated particles cannot be labeled by their coordinates in space-time, since they are indiscernibles and space-time is a classical structure with individual points.\/} In other words, we propose another structure for space-time in quantum mechanics. We suggest that entangled quantum particles define another space-time structure (quasi-metric space ${\cal S}$), quite different from that one in classical mechanics. The space or space-time coordinates in $V$ correspond to the coordinates of macroscopic and distinguishable particles or devices. The quasi-set $[x]_2$ in axiom {\bf A1} of an EPRB space-time corresponds to the `coordinates' (or corresponding quasi-metric space points) associated to the two-particle system of EPRB-correlated electrons. So, any ``nonlocal'' phenomena among EPRB-correlated particles is not nonlocal at all, since the distance between indiscernibles is always zero, according to axiom {\bf A1} in definition (\ref{xxxx}) and axiom 3 in definition (\ref{qmetric}). One interesting side result is that our system ${\cal S}$ does not allow us to localize the quantum indistinguishable particles, which are associated to the elements of $[x]_2$. That is, entangled particles do not have local coordinates in the usual sense. Nevertheless their distance to any point of $V$ is always constant.

	It is easy to verify that the proper axioms of $\langle M,d_{q}\rangle$ are consistent with those of a quasi-metric space. One natural question is: what is the role of constant $c$? Our idea is to extend the euclidian metric $d_E$ to a space which includes m-atoms $x$, i.e., Urelemente $x$ such that $m(x)$. Since $d_q(a,x)$ cannot be zero if $m(x)$ and $a$ is an ordered $n$-tuple of real numbers ($\neg(x\equiv a)$), then the simplest solution is to consider that $d_q(a,x)$ is constant. Nevertheless, if $d_q(x,a) =_E c$, then we cannot consider $V =_E \Re^n$, since in this case we can easily show that assumption $6$ in Definition (\ref{qmetric}) is not satisfied. So, constant $c$ is closely related to the open set $V$.

	Since our EPRB-space is restricted to a neighborhood $V$ of some points in $\Re^n$, it seems clear that this neighboorhood has something to do with the experimental fact that the spin measurement probe (in EPRB) is localized in a specific point of classical space-time. More specificaly, $V$ is the neighborhood of the space-time coordinates of the spin measurement probes. In other words, the region $V$ may be regarded as the set-theoretical union of two open balls in $\Re^n$ which are associated to the space-time regions occupied by the two spin measurement devices in the EPRB experiment. See FIG. 1.

	Such a mathematical model is not in conflict with not-instantaneous electromagnetic interactions between the two correlated electrons if we consider that this null distance between them allows just interactions directly related to the entangled state involved. Besides, it is a nice result that any null distance is obviously invariant under Galilean as well as Lorentzian transformations of coordinates. Despite the fact that simultaneity depends on the observer in special relativity, this simultaneity occurs between two points very close (null distance). It seems natural to admit that a relativistic approach to this kind of space-time structure is demanded. But that is a task for future works.

\section{Conclusions}

	According to FIG. 1 and our previous explanation, nonlocality in EPRB is a misunderstanding. Although the distance between the spin measurement devices is not zero, the distance between the two particles is null. So, there is no non-local phenomena at all. Besides, the distance between a given particle of the EPRB system and any point of $V$ (the space-time region occupied by the spin peasurement devices) is always $c$. This last result is a confirmation that there is no manner to localize indistinguishable particles in a specific point of space-time, but there is a way to associate these particles to neighborhoods in space-time. Such an association may cause the illusion that the particles in this system can be localized. But they cannot, since EPRB correlated particles are indistinguishable.

\newpage

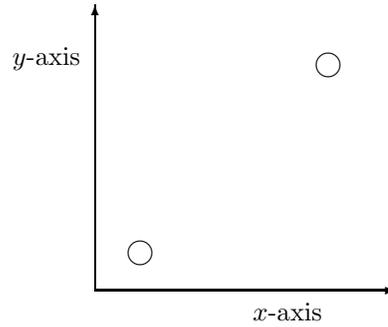
\begin{figure}[h]
\begin{picture}(42,42)
\put(14,5){\vector(1,0){40}}
\put(14,5){\vector(0,1){38}}
\put(20,10){\circle{3}}
\put(45,35){\circle{3}}
\put(35,1){$x$-axis}
\put(3,35){$y$-axis}
\end{picture}
\caption{Representation of a neighborhood $V$ of a two-par\-ti\-cle system in $\Re^2$. The two circles are two open balls in $\Re^2$ usually associated to the space regions occupied by the spin-measurement devices in the EPRB experiment. Nevertheless, we cannot say that the two-particle system is localized into this region. The region $V$ is the union of the two open balls represented by the two circles. The maximum euclidean distance between any two points of $V$ is $2c$, where $c$ is the constant in axiom {\bf A2}.}
\end{figure}

\end{document}